\documentclass[twocolumn,prb,aps,showpacs]{revtex4}

\usepackage{dcolumn}
\usepackage{amsmath}
\usepackage{epsfig}

\newlength{\figwidth}
\begin{document}


\title{X-ray Scattering and Magnetic Susceptibility Study of
doped CuGeO$_3$}

\author{Y. J. Wang}
\affiliation{Center for Materials Science and Engineering and Department of
Physics, Massachusetts Institute of Technology, Cambridge, MA 02139}
\author{Y. J. Kim}
\affiliation{Physics Department, Brookhaven National 
Laboratory, Upton, NY 11973}
\author{R. J. Christianson}
\author{S. C. LaMarra}
\author{F. C. Chou}
\affiliation{Center for Materials Science and Engineering, 
Massachusetts Institute of Technology, Cambridge, MA 02139}

\author{T. Masuda} 
\author{I. Tsukada} 
\author{K. Uchinokura}
\affiliation{Department of Applied Physics, The University of Tokyo, 7-3-1
Hongo, Bunkyo-ku, Tokyo 113-8656, Japan}

\author{R. J. Birgeneau}
\affiliation{Department of Physics, University of
Toronto, Toronto, Ontario M5S 1A7, Canada}

\date{\today}

\begin{abstract}

We report comprehensive synchrotron x-ray scattering and magnetic
susceptibility studies of the doped spin-Peierls materials
Cu$_{1-x}$Zn$_x$GeO$_3$ and CuGe$_{1-y}$Si$_{y}$O$_3$. Temperature versus
dopant concentration phase diagrams are mapped out for both Zn and Si
dopants. The phase diagrams of both Cu$_{1-x}$Zn$_x$GeO$_3$ and
CuGe$_{1-y}$Si$_{y}$O$_3$ closely resemble that of
Cu$_{1-x}$Mg$_x$GeO$_3$, including the observation that the spin gap is
established at a much higher temperature than the temperature at which the
spin-Peierls dimerization attains long-range order.  The spin-Peierls
transitions in doped samples exhibit unusual phase transition behavior,
characterized by highly rounded phase transitions, Lorentzian squared
lineshapes, and very long relaxation times. Phenomenological explanations
for these observations are given by considering the effects of competing
random bond interactions as well as random fields 
generated by the
dopants. We have also confirmed the reentrance of the spin-Peierls phase
when the temperature is lowered through the antiferromagnetic ordering
transition. The low temperature re-entrance of the spin-Peierls phase has
been explained speculatively using a local phase separation scheme between
the spin-Peierls phase and the N\'eel phase.
        
\end{abstract}


\maketitle

\section{Introduction}

Low-dimensional quantum spin systems exhibit a variety of intriguing and
often counter-intuitive properties. A prominent example of such a material
is the spin-Peierls (SP) system, which consists of an array of
one-dimensional (1D) antiferromagnetic spin-chains with S=$\frac{1}{2}$ on
a deformable three-dimensional (3D) lattice. Below the spin-Peierls
transition temperature, $T_{SP}$, the spin-chains dimerize and a gap opens
in the magnetic excitation spectrum. The discovery of an inorganic SP
compound CuGeO$_3$ made possible a systematic study of impurity effects on
SP systems. \cite{Hase1} Despite extensive experimental and theoretical
efforts devoted to the study of the effects of doping on the magnetic and
structural phase transitions in CuGeO$_3$, controversies still remain
on several fronts. One major unresolved issue is the temperature versus
concentration phase diagram for both within-chain and between-chain
dopants. 
\cite{Hase2,Coad,Sasago,Martin,Fronzes,Grenier1,Grenier2,Nakao,Masuda1,Wang1,Masuda2}
The phase diagram has shown certain
unusual characteristics as revealed by different techniques: on the one
hand, the phase behavior is so complex that there are continuing debates
about the physics underlying nearly every single part of the phase
diagram. On the other hand, there exists surprising similarities in the
phase behavior for different dopants, irrespective of the type and atomic
size of the dopant ions, the magnetic moment size, or the doping geometry.
In some theoretical models, there exists a close relation between the
spin-Peierls transition and the subsequent antiferrromagnetic phase
transition which occurs at a lower temperature in samples with a dopant
concentration $x$ below a critical value $x_c$.  Our own results suggest
that the suppression of the spin-Peierls phase as a result of doping has
no causative correlation with the emergence of the low-temperature
antiferromagnetic phase except for the fact that the dopants create 
unpaired Cu spins. The lack of direct correlation is manifested most
clearly in a comparative study of magnetic and non-magnetic ion dopings
where the high temperature suppression of the SP phase seems to follow a quite universal
behavior with different dopants while the low temperature occurrence of
the antiferromagnetic phase reflects the different magnetic properties of
the individual magnetic ion dopants.

Currently, for within-chain doping studies, both magnetic and non-magnetic
ions have been utilized, including Zn, Mg and Ni.  
\cite{Hase3,Hase2,Oseroff,Lussier,Coad,Schoeffel,Sasago,Martin,Kojima,Grenier2,Nakao,Masuda1,Wang1,Masuda2}
In early work on doped CuGeO$_3$, Zn was by far the most studied dopant. A
rich amount of information has been accumulated with a variety of
experimental techniques; however, it has turned out that
Cu$_{1-x}$Zn$_x$GeO$_3$ is far from being the ideal system, because of
concentration gradient effects and difficulties in fixing the exact Zn
concentration.\cite{Masuda1} Despite these complications, we have continued
x-ray scattering studies of Zn-doped samples in the hope that by directly
comparing our experimental results with all extant ones, we can elucidate
the underlying physics. CuGe$_{1-y}$Si$_y$O$_3$, on the other hand,
currently serves as the sole system for studying between-chain doping
effects.\cite{Oseroff,Schoeffel,Grenier1} Accordingly, we have carried out
x-ray scattering experiments on CuGe$_{1-y}$Si$_y$O$_3$ in parallel with
our Zn studies. Finally, Mg doping, regarded as the ideal dopant for
within-chain dopants, has been studied extensively by us and our results
have been published in several previous
papers.\cite{Masuda1,Wang1,Masuda2,Kiryukhin} In this paper, we will focus
on a comprehensive interpretation of the general phenomenonlogy for both
within-chain and between-chain dopings. We also will present in detail a
phenomenological model for the overall phase behavior.

Studies of the effects of dopants on the spin-Peierls transition in
CuGeO$_3$ started with the magnetic susceptibility measurements by Hase
{\it et al.} right after the discovery of CuGeO$_3$ as the first
inorganic spin-Peierls material.\cite{Hase3,Hase4} In that work, Hase {\it
et al.} reported a systematic study of the dopant concentration versus
temperature phase diagram using Zn as the dopant. Later, despite extensive
experimental and theoretical efforts devoted to trying to confirm this
phase diagram, more and more controversies seemed to turn up.
\cite{Oseroff,Schoeffel,Hase3,Hase4,Lussier,Coad,Khomskii,Fukuyama}
Scattering experiments, arguably the most direct experimental technique
for studying structural phase transitions, have provided invaluable
insights into the phenomenology of doped CuGeO$_3$. Neutron scattering
experiments have been carried out by different groups.
\cite{Hase4,Sasago,Martin,Nakao} In particular, Martin {\it et al.}
\cite{Martin} performed the first comprehensive neutron-scattering study
of Cu$_{1-x}$Zn$_x$GeO$_3$. They discovered a temperature-concentration
phase diagram in which the spin-Peierls phase transition temperature
($T_{SP}$) first decreased linearly with Zn concentration and then
appeared to show a plateau-like behavior above about 2\% Zn doping. The
plateau was presumed to persist up to more than 5\% Zn doping.
Nevertheless, significant broadening of the SP superlattice diffraction
peaks was observed in a 3.2\% Zn-doped sample which suggests that the high
doping spin-Peierls phase may not be a true independent phase. This phase
diagram was clearly in contradiction with the one obtained by Hase {\it et
al.} through susceptibility measurements.\cite{Hase3} In the original
phase diagram obtained by Hase {\it et al.}, there is a linear decrease of
$T_{SP}$ upon doping up to 2\% Zn, which is in agreement with Martin {\it
et al.}'s results. However, Hase {\it et al.} reported the disappearance
of the SP transition above 2\% Zn doping; this was indicated by the
disappearance of the characteristic dip in the magnetic susceptibility
curve, which reflects the opening of the SP spin gap and therefore
ordinarily signifies the onset of the SP transition.

At first glance, the above two phase diagrams seem to be irreconcilable
with each other. However, if one takes a closer look at the experimental
specifics, one finds that the discrepancy originates in part from the
different approaches of two experimental techniques.  Susceptibility
measurements, as a macroscopic experimental technique, are primarily
sensitive to long range order and less so to local fluctuations.
Neutron scattering, on the other hand, given sufficiently coarse
resolution and long scan times, can easily reveal the existence of
short range fluctuations.  The neutron scattering results reported by
Martin {\it et al.} strongly suggest that the high doping SP phase is
of short range order and hence not easily detected by magnetic
susceptibility measurements. Thus, care must be taken in determining
the phase diagram of the high doping phase. Similar ambiguities have
been discovered in studies of Si-doped samples.

Thus, a systematic and detailed study using both susceptibility and
scattering techniques should help to resolve some of the extant
discrepancies. In this paper, we report a detailed synchrotron x-ray
scattering and magnetic susceptibility study of the spin-Peierls
transition in both Cu$_{1-x}$Zn$_{x}$GeO$_3$ and
CuGe$_{1-y}$Si$_{y}$O$_3$. The paper is organized as follows: First we
present the results for Cu$_{1-x}$Zn$_{x}$GeO$_3$ as a prototypical
example to elaborate on the major experimental results. The general
conclusions are applied to our study of CuGe$_{1-y}$Si$_{y}$O$_3$. We then
present a phenomenological interpretation of our combined results,
in which the important roles played by competing interactions, random
fields and phase inhomogeneity will be emphasized.\cite{thesis}

In the next section, we describe the sample preparation and the
experimental techniques used in this study. In Sec.\ \ref{sec:results},
our experimental results from x-ray scattering and magnetic susceptibility
measurements are presented, including the experimentally determined phase
diagrams for both Zn- and Si-doped CuGeO$_3$. Finally, we will discuss
various theoretical and experimental issues in Sec.\ \ref{sec:discussion}.

\section{Experimental Details}
\subsection{Crystal Growth} 
 
The single crystals used in the experiment come from two sources. The
CuGe$_{1-y}$Si$_y$O$_3$ single crystals were grown at the University of
Tokyo using the traveling solvent floating zone method and preliminary
sample characterization and magnetic susceptibility results have been
published elsewhere.\cite{Masuda3} The impurity concentration of these
samples has been determined by inductively coupled plasma atomic
emission spectroscopy(ICP-AES), with an accuracy of about 0.1\%. The
Cu$_{1-x}$Zn$_x$GeO$_3$ crystals were grown by the same method at MIT.
Samples have a typical mosaic spread of less than $0.05^\circ$
full-width at half-maximum at major Bragg diffraction peaks. The
typical sample size used for both x-ray scattering and magnetic
susceptibility measurements was about $3\times3\times1$ mm$^3$. We
prepared our samples by cutting them from the end of regular growths
which have lengths over 5 cm. This procedure effectively eliminated the
gradient induced in crystal growth caused by the different doping levels
of the seed crystals.

In order to differentiate the intrinsic physics from extraneous effects
caused by simple concentration gradients, a detailed and careful
characterization of the samples is crucial. The electron probe
microscope analysis (EPMA) method was used to determine the Zn doping
concentrations. Twenty evenly spaced spots covering the whole sample
surface were used in the EPMA measurements. The variations of
concentrations were recorded as the uncertainties in the sample
concentration. The sample characterization results are summarized in
Table I. As shown in Table I, the actual Zn contents are always lower
than the nominal ones in the high doping range, which is consistent
with previously reported results. \cite{Coad,Martin} However, despite
the great disparities between these two concentrations, the
concentration gradients and fluctuations found over the surfaces remain
relatively small and comparable to those in the Mg-doped samples.  
\cite{Masuda1} Moreover, the concentration variations have similar
magnitudes for all samples, which suggests comparable effects, if any,
on the observed SP phase transitions.

\begin{table}
\caption
{Characterization of Cu$_{1-x}$Zn$_x$GeO$_3$ crystals. $T_{SP}$ and
$T_{N}$ are determined from the magnetic susceptibility measurements (see
Fig. 1).}
\label{table1}
\begin{ruledtabular}
\begin{tabular}{llll}
x (nominal)  & x (EPMA) & $T_{SP}$ (K)   & $T_{N}$ (K) \\
0.0      & $<0.003$         & 14.10(5)        &  -     \\
0.005    & $0.007(1)$  &  13.5(1)         &  -      \\
0.01      & $0.011(1)$  &  12.6(1)        &  -     \\
0.02      & $0.020(1)$   & 10.5(1)         &  2.2   \\
0.03      & $0.0235(10)$ &  9.7(2)          &  2.75 \\
0.043    & $0.0255(10)$ &  9.2(3)         &  2.9   \\
0.03      & $0.0270(15)$ & 8.8(5)    &  3.1/3.8 \\
0.046    & $0.034(3)$  &  -            &  4.3   \\
0.05      & $0.038(2)$  &  -          &  4.4    \\
0.06      & $0.038(5)$  &  -            &  4.4\\
\end{tabular}
\end{ruledtabular}
\end{table}

\subsection{Synchrotron x-ray scattering}
   
High resolution synchrotron x-ray scattering is a powerful tool in
studying structural phase transitions. High-flux synchrotron x-ray
radiation enables us to study both the critical fluctuations and the long
range order.  Our experiments were carried out at MIT-IBM beamline X20A
at the National Synchrotron Light Source. The 8.5 keV x-ray beam was
focused by a mirror, monochromatized by a pair of Ge (111) crystals,
scattered from the sample, and analyzed by a Si (111) analyzer.
Carefully cleaved samples were placed inside a Be can filled with
helium heat-exchange gas and mounted on the cold finger of a 
closed-cycle refrigerator.  The measurements were carried out in each
sample at the (1.5, 1, 1.5) SP dimerization peak position with the (H K
H) zone in the scattering plane.  The high momentum resolution of
the synchrotron x-ray beam enabled us to measure intrinsic correlation
lengths as large as 5000 \AA;  any system with a correlation length
larger than this was considered to possess long-range order (LRO).

\subsection{Magnetic susceptibility}
 
The magnetic susceptibility of the identical samples used in the x-ray
experiments has been measured using a commercial SQUID magnetometer
(Quantum Design MPMS). The samples were mounted with the $c$ axis
parallel to the applied magnetic field, and the data were taken in a
magnetic field of 500 Oe.

\section{Experimental results}
\label{sec:results}

\subsection{Magnetic susceptibility -- Zn doping}

\begin{figure}
\centerline{\epsfxsize=\figwidth\epsfbox{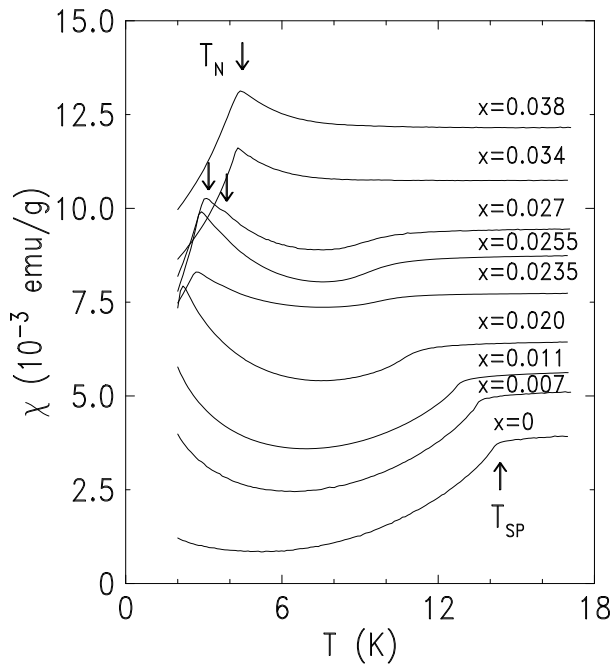}}
\caption{The magnetic susceptibility as a function of temperature for
various Zn-doped samples. Each curve is shifted by $10^{-3}$ emu/g for clarity.}
\label{fig1}
\end{figure}

Figure~\ref{fig1} shows the magnetic susceptibility as a function of
temperature for all of the Cu$_{1-x}$Zn$_x$GeO$_3$ samples. The overall
features are similar to previously reported results,
\cite{Grenier1,Grenier2,Masuda1,Masuda2} showing a rapid suppression of
the SP phase transition temperature upon doping and the appearance of a
low temperature antiferromagnetic (AF) phase. For doped samples, the
kink anomaly, which is a characteristic of the SP transition, is rounded.
We determine $T_{SP}$ in the susceptibility measurements from the
maximum of the derivative of the kink anomaly and the N\'eel
temperature, $T_{N}$, simply from the low-temperature cusp temperature.
For samples with $x$=0.0235, 0.0255, 0.34, and 0.38, a sharp
low-temperature cusp which signifies the onset of the AF state is
observed. On the other hand, the susceptibility curve of the $x$=0.027
sample shows a broad low temperature peak with two cusp positions
possibly corresponding to two N\'eel temperatures.

This is reminiscent of similar results obtained for Mg-doped samples by
Masuda and coworkers.\cite{Masuda2} Those authors carried out
susceptibility measurements on a series of Cu$_{1-x}$Mg$_x$GeO$_3$
samples, and found a double peak feature in the low temperature
susceptibility data, for $x$ in the neighborhood of 0.023. In their study,
it was argued that there exists a first order phase transition as a
function of $x$, between a dimerized antiferromagnetic phase and a uniform
antiferromagnetic phase, so that around the critical concentration $x_c
\simeq 0.023$, two AF transitions appear because of the microscopic phase
separation between these two phases. Detailed comparison of the magnetic
susceptibilities of Zn- and Mg- doped samples yield many similarities, as
expected. In Ref.\ \onlinecite{Masuda2}, the sample with the highest
Mg-doping without the broadening feature had a N\'eel transition around
2.9 K and $x_{Mg}=0.0229$. The low threshold started with $x_{Mg}=0.0237$
in which there existed a broadened cusp spanning from 3.2 K to 3.9 K. The
cusp peaked more to the low temperature side with the weight gradually
shifting to the high temperature side as the concentration was increased.
\cite{Masuda2}  We note in Fig.~\ref{fig1} similar behavior in our
Zn-doped samples, with the $x=0.0255$ Zn-doped sample showing a single
cusp at 2.9K, while the $x=0.027$ sample shows a broadened cusp over the
temperature range from 3.1 K to 3.8 K.  For $x$=0.027, the left side of the
cusp is much higher than the right side suggesting that this sample is
closer to the lower temperature threshold of the peak broadening
phenomenon.  Unfortunately, we do not have a range of doping
concentrations from $x$=0.027 to $x$=0.034 which could give us a more complete
view of the gradual shift of the weights. The important point is, however,
that we have confirmed that the regime where the susceptibility shows
significant broadening coincides with the region where, as we shall discuss in
the next subsection, the SP phase begins to lose LRO while substantial
superlattice scattering intensity is nevertheless still observable.
\cite{Wang1} Therefore, the N\'eel transition broadening must somehow be
correlated with the loss of LRO of the SP phase.

\subsection{X-ray scattering -- Zn doping}

The instrumental resolution function was experimentally measured at the
(3, 0, 0) Bragg peak, which has a Q value close to that of the (1.5, 1,
1.5) dimerization peak. The determination of the intrinsic line-shape is
complicated by the existence of the two-length scale phenomenon as has
been discussed in previous work by some of us.\cite{Wang2} Since two
superimposed profiles corresponding to two length scales were always
present in the critical scattering in the close vicinity of the SP
transition for $x \leq 0.02$ samples, a unique determination of the
intrinsic cross-section was difficult.  However, only a single scattering
profile was observable in the critical scattering for $x \geq 0.02$
samples, where dopant effects play a more important role than the near
surface dislocation effects which are presumably responsible for the large
length scale fluctuations in samples with LRO. For typical second order
phase transitions, the critical scattering cross section assumes a
Lorentzian lineshape. Any deviation from this line-shape can provide
information on the effects of disorder on the underlying phase transition.

\begin{figure}
\centerline{\epsfxsize=\figwidth\epsfbox{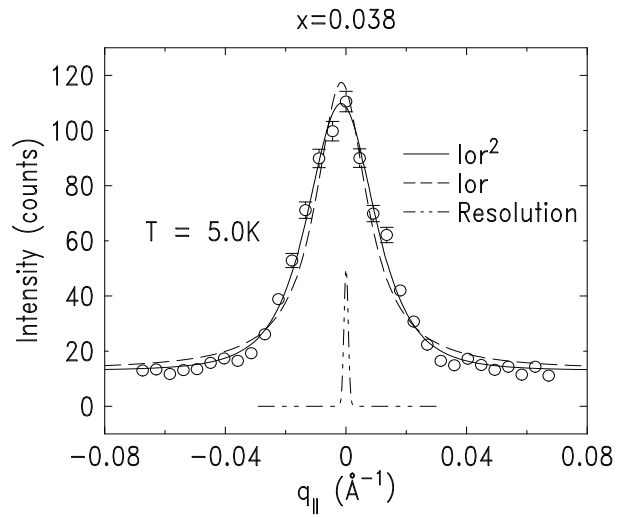}}
\caption{Longitudinal scan profile of the
superlattice peak, (1.5, 1, 1.5), at $T$=5 K for $x$=0.038.
The solid and dashed line shows the correponding results fitted by
Lorentzian and Lorentzian squared lineshapes convoluted with the
instrumental resolution function.}
\label{fig2}
\end{figure}

In Fig.~\ref{fig2}, we show the low temperature (1.5, 1, 1.5)
superlattice peak profile for the $x$=0.038 sample.  It is evident that
the resolution function is much narrower than the intrinsic cross
section. Both Lorentzian and Lorentzian squared cross sections have
been used to fit the data.  The Lorentzian squared lineshape gives a
discernably better fit than a simple Lorentzian lineshape.

\begin{figure}
\centerline{\epsfxsize=\figwidth\epsfbox{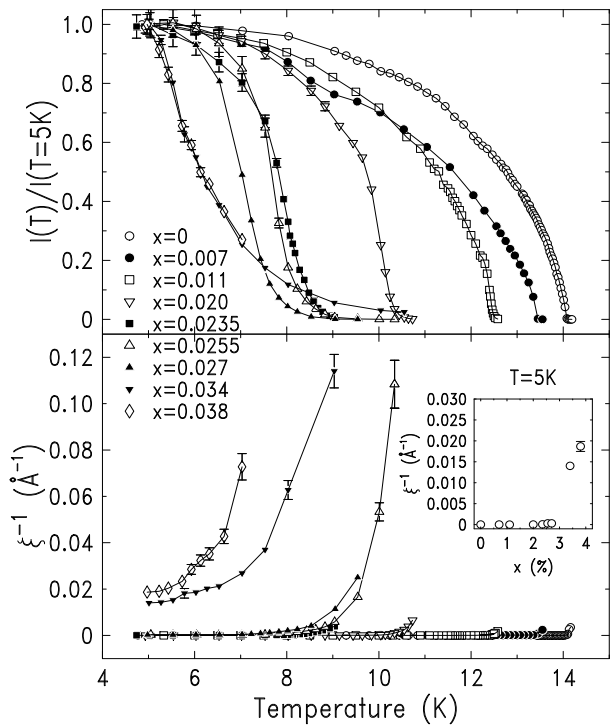}}
\caption
{The (1.5, 1, 1.5) SP peak intensity (a) and the corresponding 
inverse correlation length (b) as functions of temperature for various
Zn-doping concentrations. The inset shows the impurity-concentration   
dependence of the low-temperature SP inverse correlation length}
\label{fig3}
\end{figure}

Figure~\ref{fig3}(a) shows the temperature dependence of the (1.5, 1, 1.5)
dimerization peak intensity for various Cu$_{1-x}$Zn$_x$GeO$_3$ samples.
We normalize the intensity of each sample data set using the dimerization
peak intensity at 5 K. As is obvious in the figure, in sharp contrast to
the transition in pure CuGeO$_3$, the SP transitions for the doped samples
are noticeably rounded, and further the rounding increases progressively
with increasing dopant concentration.  We believe that this substantial
rounding is intrinsic and cannot simply be accounted for by local
concentration gradients for the following reasons: (i) The impurity
concentration is too low to create any substantial correlated effect.  
\cite{Birgeneau2} (ii) There exists a monotonic increase of the rounding
in the order parameter as the doping concentration is increased, while
experimentally the measured concentration gradients are of similar
magnitude for all of the samples studied. For example, the $x$=0.038 sample
has a concentration variance of less than 0.2\%, from Table I. However,
that sample's SP dimerization peak intensity increases sufficiently
gradually as the temperature is lowered that it does not achieve
saturation for the whole temperature range studied. In sharp contrast, no
significant rounding was found in the $x$=0.011 sample which has a similar
concentration variance.

Not only does the order parameter temperature dependence change
dramatically upon doping, the corresponding saturation correlation length
also behaves quite differently from that in the undoped sample. Figure~\ref{fig3}(b) shows the temperature dependence of the longitudinal inverse
correlation length for the different Cu$_{1-x}$Zn$_x$GeO$_3$ samples. The
minimum inverse correlation length which the sample achieves as a function
of $x$ is plotted in the inset. Unlike the pure material, in which very
near $T_{SP}$ the inverse correlation length decreases linearly to zero as
$T_{SP}$ is approached from above because of the second length scale
critical fluctuations, the inverse correlation length of the doped samples
appears to approach zero in an asymptotic way without a clear signature of
the exact temperature at which it becomes zero.  This complication makes
an unequivocal determination of $T_{SP}$ difficult. Within our
experimental accuracy, samples attain LRO (that is, the correlation length
exceeds 5000 \AA) only for $x \leq 0.027$. Substantial intensity is still
observable at the superlattice peak positions for the $x$=0.034 and $x$=0.038
samples which have $x>x_c \equiv 0.027$, although the superlattice peaks
are not resolution-limited in width. The critical dopant concentration,
$x_c$, which we have determined for Cu$_{1-x}$Zn$_x$GeO$_3$ is slightly
higher than that for the Cu$_{1-x}$Mg$_x$GeO$_3$ samples. However, this
could be due in part to the different techniques used in determining the
concentrations.

\begin{figure}
\centerline{\epsfxsize=\figwidth\epsfbox{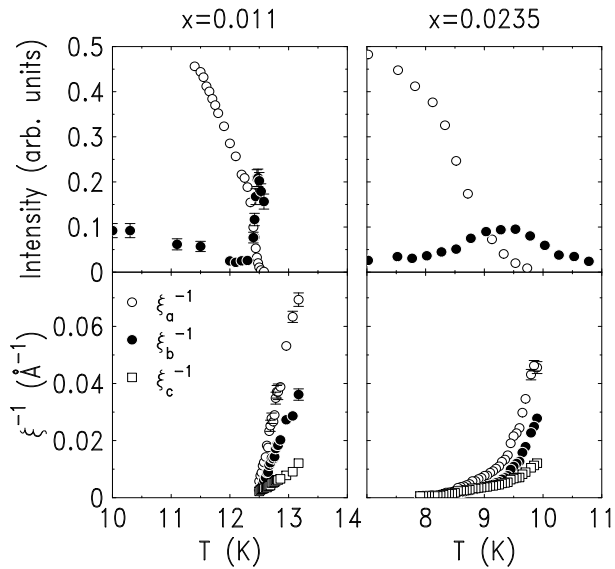}}
\caption{Top: Peak intensity
of the superlattice peak (empty symbols) and the diffuse scattering intensity 
at (1.5 1.03 1.5), plotted as a function of temperature.
Bottom: Inverse correlation length  of the
critical scattering along the a, b, and c directions. }
\label{fig4}
\end{figure}

In Fig.~\ref{fig4}, we show the results from the fits for the peak
intensity, which for LRO is the order parameter squared, and the inverse
correlation length, $\xi^{-1}$ for samples with $x$=0.011 and $x$=0.0235.  
Also plotted is the diffuse scattering intensity at the wing ($\delta q=0.022$ \AA$^{-1}$) of the 
dimerization superlattice peak, which is a measure of the amplitude of the critical
fluctuations. The transition is
slightly broadened as compared to the SP transition in pure CuGeO$_3$.
This is reflected both in the smearing of the order parameter onset
temperature and the temperature range of the diffuse scattering.
The inverse correlation length follows more or less a linear relationship
with temperature.  For the samples with $x$=0.0235, the broadening in the
temperature dependence of both the order parameter and the diffuse
scattering has increased significantly. The corresponding inverse
correlation length first exhibits a linear relationship as a function of
temperature with decreasing temperature.  However, close to the
transition, $T \leq 9.5$ K, the rate of the divergence slows down and
$\xi^{-1}$ appears to approach zero asymptotically, that is, with zero
slope.  It is interesting to note that if the linear behavior is
extrapolated to intersect the temperature axis, one obtains a temperature
which roughly matches that of the maximum of the diffuse scattering.

SP transitions in doped samples have shown unconventional
characteristics that are absent in typical second-order phase transitions.
One of these is that when the doped system undergoes a SP transition, the
times for the system to reach phase equilibrium are anomalously long. In
x-ray scattering experiments, this is reflected in a continuous change in
the peak intensity and width as a function of time. This anomaly manifests
itself in pronounced thermal hysteresis in the order parameter
measurements if the system is not kept sufficiently long at each
temperature. This slow dynamics behavior is
not observed in pure CuGeO$_3$. Under the same experimental conditions,
upon rapid change of temperature by several degrees, the pure system
achieves equilibrium within minutes, in sharp contrast to the highly doped
samples. We characterized the slow dynamics by monitoring the intensity of
the superlattice peak after a sudden change in temperature as a function
of time while keeping the final temperature constant. The time scans for
the system in equilibrium show an oscillating behavior with the centerline
remaining flat. The oscillations reflects the temperature fluctuations at
the sample place and agree well with the temperature fluctuation patterns
recorded from the temperature controller.

\begin{figure}
\centerline{\epsfxsize=\figwidth\epsfbox{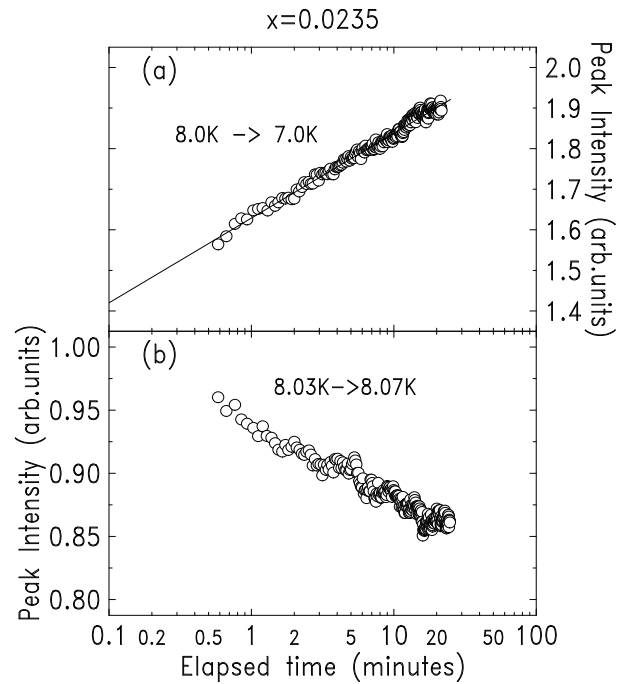}}
\caption{(a) Semi-logarithmic plot of the peak intensity of the superlattice 
peak as function of time after a sudden temperature change from 8.0 K to 
7.0 K for
$x$=0.0235. (b) Similar plot after a temperature change from 8.03 K to 8.07
K.}
\label{fig5}
\end{figure}

Figure~\ref{fig5} shows representative time scans for $x$=0.0235 for both
heating and cooling temperature runs on a semi-logarithmic scale.  After the sample was quenched from
8K to 7K, it was observed that the peak intensity increased monotonically
with time and did not saturate after more than half an hour. There are, 
in addition, oscillations identical to those observed
in the pure sample. This reveals that the slow dynamics are not caused by
poor thermal contacts. The slow dynamics behavior exists even upon change
of temperature by an amount as small as 0.05K as shown in 
Fig.~\ref{fig5}(b).
The linear relationship in Fig.~\ref{fig5} suggests that the peak intensities are following
time logarithmic behavior.  In addition to the anomalous behavior of the
peak intensity, the peak width also changes continuously. A slow increase
of intensity is accompanied by a slow decrease in peak width and {\it
vice versa}. This means that the time dependence originates in slow growth
or contraction of the domains after a sudden change in temperature.

To understand the origin of the slow dynamics, we first recall that for SP
systems, the phase transition occurs as the neighboring spins pair up and
create a local gap in the excitation spectrum, and eventually LRO is
achieved when the 3D phase coherence of these local
dimerizations is established. In pure CuGeO$_3$, these two events happen
at the same temperature.  For doped samples, on the other hand, we
observed substantial superlattice intensity even when the system had only
SRO, which suggested that SP order had been established locally though the
system had difficulty in achieving global phase coherence.  Similar
behavior has been observed in random field Ising model (RFIM)
systems.\cite{Feng} The slow dynamics, then, results from the expansion or
contraction of the SRO region as a function of time. However, we do not
have quantitative information on the time evolution of the peak width.

\subsection{X-ray scattering -- Si doping}

As discussed in Sec. I, one of the findings in the studies of 
doped CuGeO$_3$
is the close resemblance of the phase diagram for
both within-chain and between-chain dopings.
\cite{Grenier1,Grenier2,Masuda1,Masuda2,Renard,Weiden} That is, substitution
on the Cu site and substitution on the Ge site have yielded very similar
effects on the SP phase, including rapid suppression of the SP transition
temperature, occurrence of a low-temperature AF phase, and the concomitant
co-existence of the SP and AF phases. 

In this section, we describe synchrotron x-ray diffraction experiments on
CuGe$_{1-y}$Si$_y$O$_3$ samples. As expected, most of our experimental
observations are similar to the Zn-doping results presented in the
previous section. However, we find that Si-doping is two to three times as
effective in destroying the SP phase as compared to Zn- or Mg- doping, in
agreement with earlier workers.\cite{Grenier2}

\begin{figure}
\centerline{\epsfxsize=\figwidth\epsfbox{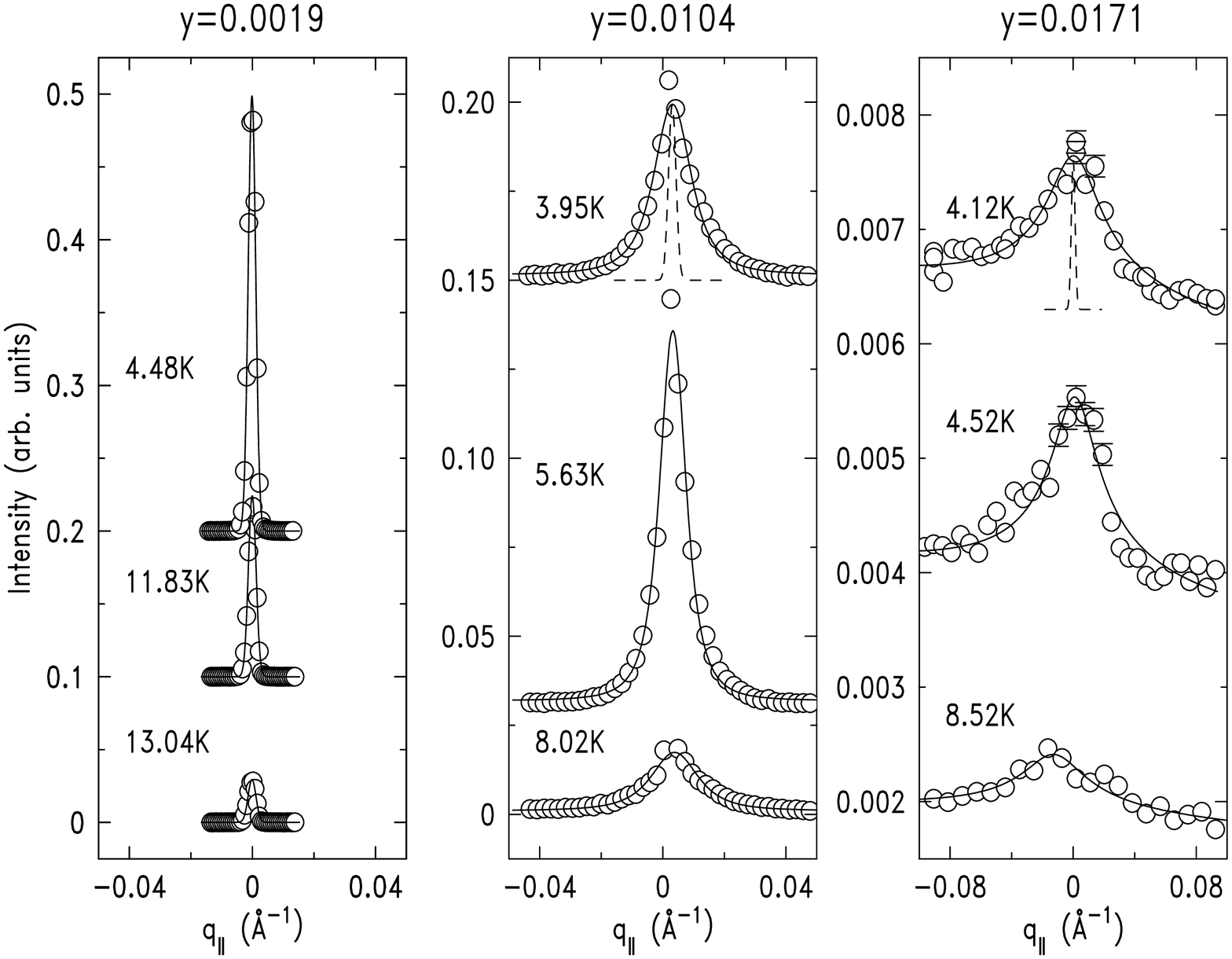}}
\caption{Representative longitudinal scans at the (1.5, 1, 1.5) SP
dimerization peak position. The solid lines are the results of fits to
a Lorentzian squared lineshape as described in the text.
The instrumental resolution function is shown as dashed lines. }
\label{fig6}
\end{figure}

A representative set of longitudinal (parallel to the scattering vector)
x-ray scans around the (1.5, 1, 1.5) SP dimerization peak position for the
$y$=0.0019, 0.0104, and 0.0171 samples at various temperature is shown in
Fig.~\ref{fig6}. Similar to the Zn-doping case, the low-temperature SP
peak width is resolution limited in the $y$=0.0019 sample, indicating that
the SP phase possesses LRO. On the other hand, the SP peaks in the
$y$=0.0104 and $y$=0.0171 samples are broadened at all experimentally
accessible temperatures, and therefore only SRO is present. To extract the
correlation length, the data for all samples were fit to a two-dimensional
convolution (longitudinal and transverse directions) of the experimentally
measured Gaussian resolution function with the intrinsic cross section
taken as Lorentzian squared.

In Fig.~\ref{fig6}, the peak intensity and the associated inverse
correlation length obtained from the fitting are shown as functions of
temperature. For the $y$=0.0019 sample, with less than 0.2\% of Si
substitution, the SP transition temperature has been reduced by 1 K.
However, other than that, the $x$=0.0019 sample behaves analogously to pure
CuGeO$_3$. The effects of doping can be more obviously seen as we
gradually increase the dopant concentrations. For the samples with higher
doping levels, smearing of the order parameter onset temperature, which is
reminiscent of the behavior in Cu$_{1-x}$Zn$_x$GeO$_3$ and 
Cu$_{1-x}$Mg$_x$GeO$_3$, can be identified. In
addition, over the whole temperature range studied, the $y$=0.0104 and the
$y$=0.0171 samples never achieve LRO and the correlation lengths at low
temperatures saturate at the finite values of about 300 \AA\ and
90 \AA, respectively. Moreover, for the $y$=0.0104 sample, this saturation of
the correlation length at low temperatures is preempted by the low
temperature re-entrance of the SP phase, which is characterized by a
suppression of the dimerization peak intensity and a concomitant decrease
in the correlation length. The re-entrance, as already elaborated in our
Mg doping studies,\cite{Kiryukhin} can be understood as arising from local
phase separation of the SP and AF micro-regions. For the $y$=0.0171 sample,
despite the large error bars due to the low intensity, the reentrance at
around 4 K is still observable from the drop in the peak intensity upon
cooling from 4.5 K to 4.1 K. This temperature range also includes the
N\'eel temperature of this sample.

\begin{figure} 
\centerline{\epsfxsize=\figwidth\epsfbox{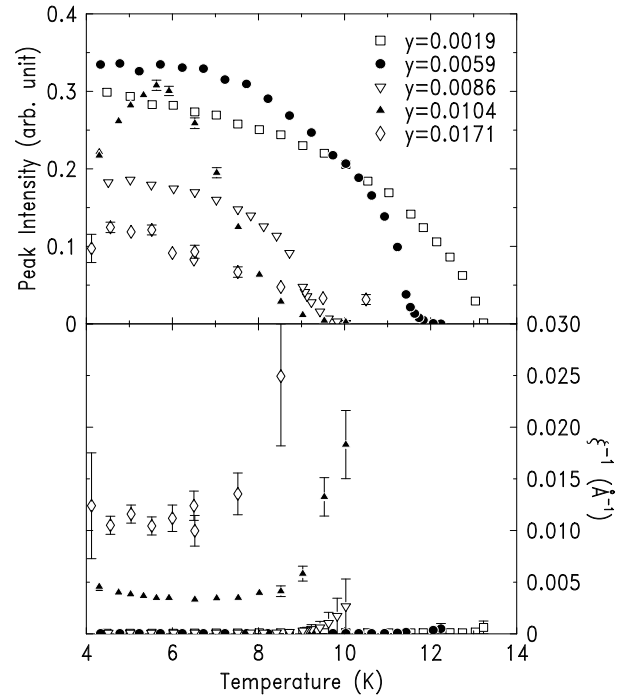}}
\caption{Peak intensity (top) and inverse correlation length (bottom) of the
SP peak as a function of temperature for
the CuGe$_{1-y}$Si$_y$O$_3$ samples.}
\label{fig7}
\end{figure}

\subsection{Phase diagram}

In order to construct comprehensive temperature versus impurity
concentration phase diagrams from our x-ray scattering and susceptibility
measurements, we define two characteristic temperatures, both related to
the SP transition but defined by the physics on different length-scales.
The higher transition temperature $T_m$, is defined as the temperature
corresponding to the peak temperature of the critical fluctuations as
measured from diffuse scattering intensity. We defined the 
counterpart
of $T_m$ in the bulk susceptibility measurements as the temperature at
which the derivative of the susceptibility ($d\chi/dT$)  reaches a
maximum. The lower transition temperature T$_s$ is defined as the
temperature at which the lattice dimerization achieved LRO, that is, when
the inverse correlation length reaches zero, which in our case means
$\xi^{-1} < 0.0002$ \AA$^{-1}$.

\begin{figure}
\centerline{\epsfxsize=\figwidth\epsfbox{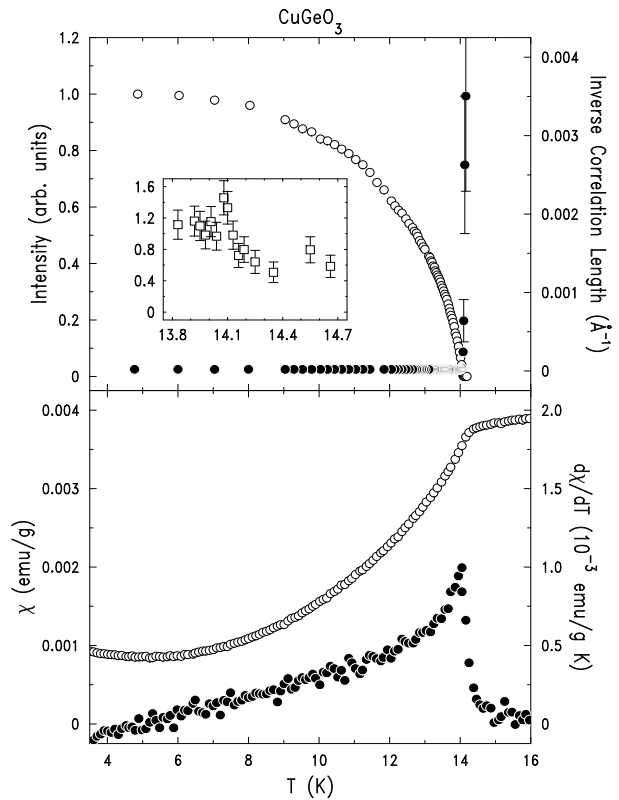}}
\caption{Top: Temperature dependence of the peak intensity (empty) and
the inverse correlation length (filled) of CuGeO$_3$. Inset: Temperature
dependence of diffuse scattering intensity at $\delta q=2.1 \times 
10^{-3}$ $\AA^{-1}$. Bottom:
Temperature dependence of the magnetic
susceptibility (empty circles) and its temperature derivative (filled 
circles).} 
\label{fig8}
\end{figure}

For the $x$=0 data shown in Fig.~\ref{fig8}, there is no ambiguity in the
definition of $T_{SP}$. The superlattice peak intensity vanishes above
$T$=14.1 K while the corresponding correlation length diverges precipitously
at the same temperature. In the inset, we show the diffuse scattering intensity at the wing
(${\delta}q=2.1\times10^{-3}\AA^{-1}$) of the dimerization peak.
It is clear that
the critical fluctuation amplitude reaches a maximum at 14.1 K as well.  
The susceptibility measurement shown in the bottom panel of Fig.~\ref{fig8} also exhibits a sharp kink anomaly at 14.1 K, at which
temperature the derivative shows a sharp peak. From these observations, we
find $T_{SP}=T_m=T_s=14.1$ K for $x$=0. Physically this means that the
temperature at which the spin gap appears coincides with the temperature
at which long-range coherence of the dimerization is achieved. This
perfect agreement by two different techniques, one magnetic and one
structural, proves that the SP transition is a well-defined second order
phase transition and it also demonstrates the equivalence of the two
definitions for the transition temperature. However, as discussed in our
previous papers and as is evident from the data presented in this paper, a
discrepancy develops between these two definitions as one progresses into
the doped regime. It is then difficult to define $T_{SP}$ unambiguously.

\begin{figure}
\centerline{\epsfxsize=\figwidth\epsfbox{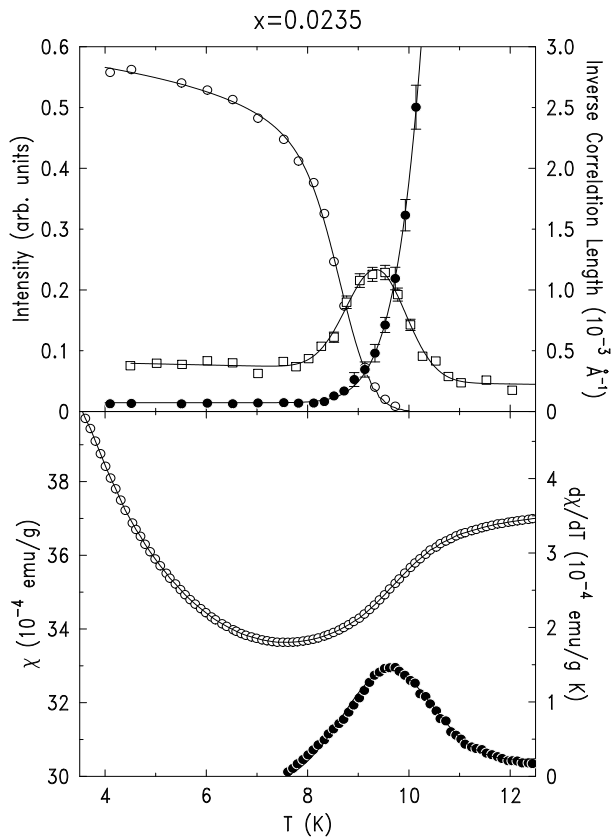}}
\caption{Top: Temperature dependence of the peak intensity (empty) and
the inverse correlation length (filled) of 
Cu$_{0.9765}$Zn$_{0.0235}$GeO$_3$. Also plotted in
square symbols is the temperature dependence of the diffuse scattering
intensity at $\delta q=2.2 \times
10^{-2}$ \AA$^{-1}$. Bottom:
Temperature dependence of magnetic
susceptibility (empty) and its temperature derivative (filled).} 
\label{fig9}
\end{figure}

In Fig.~\ref{fig9}, we show the experimental results for the $x$=0.0235
sample. In the upper panel where x-ray measurements of the dimerization
peak are presented, we find smearing of the onset temperature of the order
parameter, with substantial intensity starting to appear around 10 K while
the critical fluctuation amplitude reaches a maximum around 9.3 K. At the
same time, the correlation length does not diverge until the temperature
reaches $T_s \approx 7$ K on cooling. In the bottom panel, magnetic
susceptibility measurements are shown on the same temperature scale. The
kink anomaly corresponding to the SP transition is rounded
and the derivative of the susceptibility shows a broad peak around $T_m
\approx 9.5$ K. If a linear extrapolation method, which was adopted in
Ref.\ \onlinecite{Masuda2}, is used, we obtain$ T_{SP} \sim 10$ K.  
These seemingly contradictory results partially explain the difference
among the varied phase diagrams reported in the literature.

\begin{figure}
\centerline{\epsfxsize=\figwidth\epsfbox{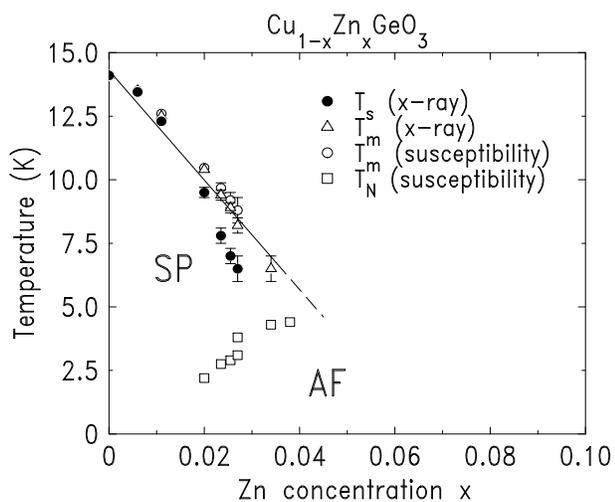}}
\caption{Experimentally determined $T-x$ phase diagram of
Cu$_{1-x}$Zn$_x$GeO$_3$.}
\label{fig10}
\end{figure}

In Fig.~\ref{fig10}, we plot the $T-x$ phase diagram for 
Cu$_{1-x}$Zn$_x$GeO$_3$ using the above
definitions from both x-ray and susceptibility results. Our phase diagram
is in several ways distinctive: The phase boundaries determined by
different definitions of $T_{SP}$ do not match each other except for $x$=0,
with the divergence increasing as the doping level increases.  The phase
boundary determined from $T_s$ is nearly vertical around $x$=0.027. No LRO
can be attained above this concentration level, although substantial
intensity at the dimerization peak position is still observed. Note that
the AF transition in the susceptibility measurement at $x$=0.027 is also
broadened. On the other hand, the phase boundary determined from $T_m$
follows an approximately linear variation with the Zn concentration. In
particular, for $x$=0.034, while $T_s$ cannot be defined due to the lack of
LRO, one can still determine $T_m$ because there still exists substantial
intensity at the dimerization peak position in the critical temperature
regime.

\begin{figure}
\centerline{\epsfxsize=\figwidth\epsfbox{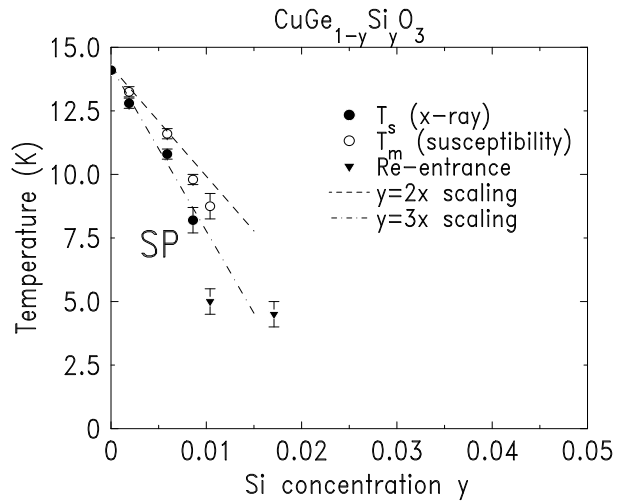}}
\caption{Experimentally determined T-y phase diagram of
CuGe$_{1-y}$Si$_y$O$_3$. The solid line shown in Fig.~\ref{fig10} is rescaled 
to illustrate the $y=2x$ and $y=3x$ scaling.}
\label{fig11}
\end{figure}

In Fig.~\ref{fig11}, we also present the T-y phase diagram for
CuGe$_{1-y}$Si$_y$O$_3$. \cite{Katano2} The same definitions for $T_s$ and
$T_m$ are used. The phase diagram is very similar to those of
Cu$_{1-x}$Mg$_x$GeO$_3$ and Cu$_{1-x}$Zn$_x$GeO$_3$, with a linear
suppression of the SP phase upon doping and a critical concentration
around $x_c$=0.009, above which no LRO can be observed. The value of $x_c
\approx 0.009$ is slightly less than half that of the critical 
concentration
$x_c$=0.021 and $x_c$=0.027 in the Mg- and in Zn-doped cases,
respectively. Also, one should note that the slope of the linear line in
Fig.~\ref{fig11} is two to three times that in Fig.~\ref{fig10}, which
suggests that the Si-doping is more than twice as effective in destroying
the SP phase as the within-chain Mg and Zn dopants. Previous workers have
rationalized this by noting that the Si-dopant affects more than one
chain.\cite{Grenier2} In addition, as we shall discuss in the next
section, the Si will create a random elastic field on the Cu site which in
a pseudo-spin description will act like a random field which is also
destructive of the order.

\section{Discussion}
\label{sec:discussion}

The first conundrum in the diluted CuGeO$_3$ problem is why the SP phase
is destroyed by only $\sim 2$\% Zn or Mg and even less in the Si case. In
comparison, a much higher doping concentration is required to destroy the
Haldane gapped state in the $S=1$ spin chain.\cite{Ajiro}  Khomskii {\it et
al.} \cite{Khomskii} first proposed a soliton model as a starting point
for the study of the doped CuGeO$_3$ problem. The soliton picture is
consistent with both numerical simulations\cite{Ain,Arai,Dobry1,Dobry2}
and experiments. \cite{Fagot-Revurat,Horvatic} However, Khomskii and
coworkers note that if diluted CuGeO$_3$ is treated in a simple
percolation picture, then the soliton model naturally yields a critical
concentration $x_c \sim10\%$. Here $x_c$ is determined by the
concentration at which the average dopant spacing equals the soliton
width. This implies that there must be a separate mechanism that is
responsible for the rapid destruction of the SP phase.

\begin{figure}
\centerline{\epsfxsize=\figwidth\epsfbox{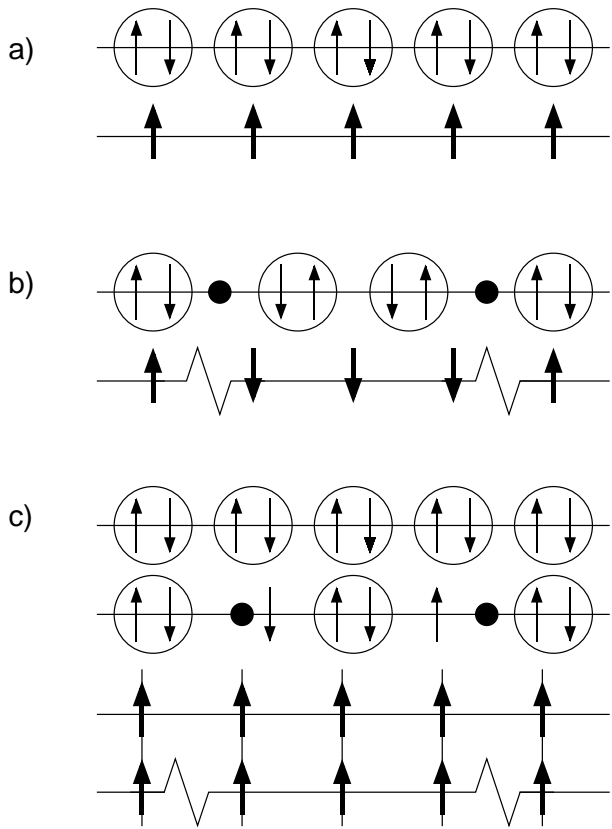}}
\caption{(a) Pseudo-Ising model mapping for single chain
without dopant and (b) with dopant (c) Pseudo-Ising model with interchain 
interchain}
\label{fig12}
\end{figure}

In order to discuss this, we follow Harris {\it et al.} \cite{Harris} and map the dimerized SP system
onto an effective 3D Ising model, in which the two dimer configurations possible in a given chain are
associated with the up and down states of the Ising pseudo-spins as shown in Fig.~\ref{fig12}(a).
When a few Cu ions are replaced by Zn ions in an isolated chain, it is energetically favorable for
the Ising pseudo-spin to change sign across the Zn dopant. In Fig.~\ref{fig12}(b), we show that the
dopant effectively creates an antiferromagnetic bond in the ferromagnetic pseudo-Ising model.  Note
that the dopants can cut the chains into segments with either an even or an odd number of spins in
them.  The ``even" segments favors the dimer configuration in which all the spins are paired, since
this configuration has lower energy in comparison with the other one, which has two unpaired spins at
the ends of the segments. For the ``odd" segment, on the other hand, the two dimer configurations are
energetically equivalent. According to numerical simulations,\cite{Hansen} however, the soliton in
the odd segments prefers to be in the center of the segment, thereby creating two smaller segments
with opposite pseudo-Ising spins.

As temperature is lowered, a uniformly ordered phase, as shown in 
Fig.~\ref{fig12}(c) is preferred, since the interchain interaction prefers to
keep all the dimers in phase in order to achieve 3D LRO. Therefore,
depending on the energetics of the interchain interaction and the soliton
creation energy, locally ordered dimers with the ``wrong" pseudospin may be
flipped at low temperatures. We note a similar mapping by Mostovoy {\it et
al.}, \cite{Mostovoy1} in which the equivalent fields were taken to
infinity to solve the model.

The essential physics thus involves a competition between the interchain 
interaction which favors a uniform phase of the dimerization and the local 
soliton energy which is minimized by a change in phase of the dimerization 
at the dopant site. It is this competition between intrachain and 
interchain 
energies which leads to the fragility of the LRO spin-Peierls state.

Another system of this kind is a magnetic system with both ferro- and
antiferromanetic bonds. For this system, the ordering problem with
competing interactions has been well addressed. \cite{Aeppli1,Aeppli2} In
particular, Aeppli, Maletta and coworkers proposed a phenomenological
cluster spin-glass model
\cite{Aeppli1,Aeppli2,Maletta1,Maletta2,Reich1,Reich2} to explain the
evolution of the ground state from ferromagnetic LRO to spin-glass.  

We should emphasize that our observation in doped CuGeO$_3$ of a
Lorentzian squared cross section instead of Lorentzian is significant. By
way of contrast, in a simple dilution induced percolation problem,
\cite{Birgeneau2} even above the percolation threshold, the diffuse
scattering should always be Lorentzian, which is characteristic for
fluctuations of thermal or geometrical origin only.  The occurrence of a
Lorentzian squared lineshape implies competing interaction or random field
dominated physics.\cite{Birgeneau1} In the Si-doped case, because of the
low symmetry of the Ge site the dopant will generate both competing
interchain interactions and random fields. Microscopic calculations are
required to determine which are more important quantitatively.

In addition to the unusual SP transition, the AF transitions in doped
CuGeO$_3$ samples also exhibit abnormal characteristics. Neutron
scattering experiment on 3.4\% (nominal) Zn-doped CuGeO$_3$ \cite{Hase4}
revealed that the AF order parameter did not saturate even at 1.4 K. In
addition, the transition itself was considerably rounded. Fitting the
order parameter using a Gaussian distribution of the transition
temperature, Hase {\it et al.} obtained $\beta = 0.22\pm 0.02$ which did not fit
into any extant 3D universality class. In the $\mu$SR studies,
\cite{Tch,Garcia} the N\'eel ordering process was described as an
inhomogeneous process with islands of freezing spins emerging from the
paramagnetic phase. Although the transition was mistakenly interpreted in
a spin glass picture, it revealed that the N\'eel transition was not a
homogenous process despite the fact that it yields a 3D LRO magnetic
state.  We believe that the inhomogeneous ordering of the AF transition is
due to the coexistence of an inhomogeneous SP phase and the AF phase
involving local phase separation.

In order to understand the competition between the SP and AF phases, it is
illuminating to compare Cu$_{1-x}$Zn$_x$GeO$_3$ with
Cu$_{1-x}$Ni$_x$GeO$_3$. In the case of Ni-doping case,\cite{Grenier1} due
to the non-zero magnetic moment of the Ni ion, the AF ordering takes place
at a quite different temperature from that of Zn- or Mg- doping, while the
effect on the SP phase boundary seems to be the same. This experimental
observation suggests that the SP phase transition and the AF transition
should be treated independently, that is, the suppression of the SP
ordering can not be simply attributed to the competition from the AF
ordering. Although these two phases are intricately correlated as
evidenced by the re-entrance of the SP phase upon the onset of the AF
phase, the onset of the AF ordering is only indirectly related to the
destruction of the SP phase.

As discussed in detail in the section on Si-doping and in Ref.~\onlinecite{Kiryukhin}, we observed at
approximately the same temperature as the onset of the AF ordering, that
the SP phase underwent a reentrant transition which was evidenced by the
increase in the dimerization peak width upon cooling through N\'eel
temperature. Additionally, the reentrant process was accompanied by slow
dynamics and hysteresis. The reentrance and the hysteresis are unusual in
the sense that they coincide with the onset of the AF ordering over the
doping range studied, which implies that the reentrance is directly
related to the AF ordering.

\begin{figure}
\centerline{\epsfxsize=\figwidth\epsfbox{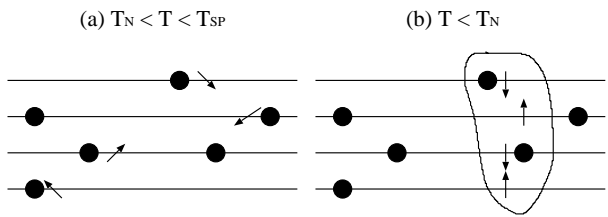}}
\caption{(a)The free spin configuration above the AF transition and (b)  
Antiferromagnetically ordered
islands below the AF 
transition.}
\label{fig13} 
\end{figure}

As shown in Fig.~\ref{fig13}, we propose a local phase separation
scenario to explain this unusual AF transition and the coexistence of the
AF and SP phases. As shown in Fig.~\ref{fig12}, solitons bear double
identities in the system, they involve both structural anti-phase domain
walls for the SP phase and the free spins for the underlying magnetic
system.  For materials with $x < x_c$ or $y < y_c$, the temperature is
decreased below the SP transition temperature, the interchain interactions
lead to the creation of a LRO SP phase. For even segments, two solitons
are created at the end -- see Fig.~\ref{fig12}(c), while for odd
segments, the solitons that were located at the center will move to the
end of the segment to bind to the dopants. As a result, a LRO SP order in
fact creates many solitons that bind to the dopant sites. These free spins
have antiferromagnetic correlations among them, and the dimerized spins
will be partially polarized to mediate the magnetic interactions. At the
N\'eel temperature, the spins eventually develop AF order. We believe that
the key factor in inducing the reentrance is the non-uniform nature of the
AF state. Above the AF transition, the free spins are disordered. However,
as the system is cooled, the antiferromagnetic interaction energy will
become relevant. The free spins are actually mobile and can move freely in
each segments. However, due to the confining potential of the interchain
interations the solitons become bound to the dopants. The onset of the AF
state, however, will compete with this configuration because, the lowest
energy state will be achieved by moving all these solitons together, in
other words, phase separation into soliton-rich and soliton-poor regions,
which will inevitably disrupt the originally established SP order.

The final state is the result of this competition between the SP order
and AF order. Since the solitons are bound by the dopants, the phase
separation can happen only locally. It is not currently known whether the
resulting state involves islands or stripes or some other geometry.
Further theoretical work must be done in this direction. Clearly, however,
the slow dynamics and re-entrance can naturally be explained in this way.

\begin{acknowledgements}

We thank G. Shirane and V. Kiryukhin for valuable discussions, and Doon Gibbs for allowing us to use
the 4K close-cycle refrigerator. The work at MIT was supported by the NSF-LTP Program under Grant No.
DMR-0071256. Work at the University of Toronto was supported by the Natural Science and Engineering
Research Council of Canada. The work at the National Synchrotron Light Source, Brookhaven National
Laboratory, was supported by the U.S. Department of Energy, Division of Materials Sciences and
Division of Chemical Sciences, under Contract No. DE-AC02-98CH10886. Research at the University of
Tokyo was partially supported by Grant-in-Aid for COE Research ``SCP Project" from the Ministry of
Education, Sports, Science and Technology of Japan.

\end{acknowledgements}

\end{document}